\documentclass[journal,twoside,web]{ieeecolor}
\usepackage{jsen}
\usepackage{cite}
\usepackage{amsmath,amssymb,amsfonts}
\usepackage{algorithmic}
\usepackage{graphicx}
\usepackage{textcomp}
\usepackage{wrapfig}
\usepackage{rotating}
\usepackage{color}
\usepackage{longtable}
\usepackage{array,makecell}
\usepackage{verbatim} 
\usepackage{fancyhdr}
\usepackage{algorithm}

\usepackage{multirow}
\usepackage{subfig}
\usepackage{makecell}
\usepackage{tabularx}
\def\BibTeX{{\rm B\kern-.05em{\sc i\kern-.025em b}\kern-.08em
    T\kern-.1667em\lower.7ex\hbox{E}\kern-.125emX}}
\definecolor{abstractbg}{rgb}{0.89804,0.94510,0.83137}
\begin{document}
\title{\vspace{0.5cm}\LARGE \bf
A Deep Learning Sequential Decoder for Transient High-Density Electromyography in Hand Gesture Recognition Using Subject-Embedded Transfer Learning}

\author{Golara Ahmadi Azar{$^{*}$}, Qin Hu{$^{*}$}, Melika Emami, Alyson Fletcher, Sundeep Rangan, S. Farokh Atashzar{$^{**}$}
		\thanks{
		This material is based upon work supported by the US National Science Foundation under Grant No \#2121391.\\
  Azar, Emami, and Fletcher are with the Department of Electrical and Computer Engineering, University of California Los Angeles (UCLA), Los Angeles, CA, 90095 USA. Fletcher is also with the Department of Statistics, Mathematics, and Computer Science, UCLA. Hu, Rangan, and Atashzar are with the Department of Electrical and Computer Engineering, New York University (NYU), New York, NY, 11201 USA. Rangan is also the Director of NYU WIRELESS. Atashzar is also with the Department of Mechanical and Aerospace Engineering, Biomedical Engineering, NYU WIRELESS, and NYU Center for Urban Science and Progress (CUSP).  \\
		{$^{*}$} Azar and Hu share the first authorship.\\
		{$^{**}$} Corresponding author: {\tt\footnotesize f.atashzar@nyu.edu}.}
	    }

\IEEEtitleabstractindextext{%
\fcolorbox{abstractbg}{abstractbg}{%
\begin{minipage}{\textwidth}%
\begin{wrapfigure}[14]{r}{4.4in}%
\includegraphics[width=4.3in]{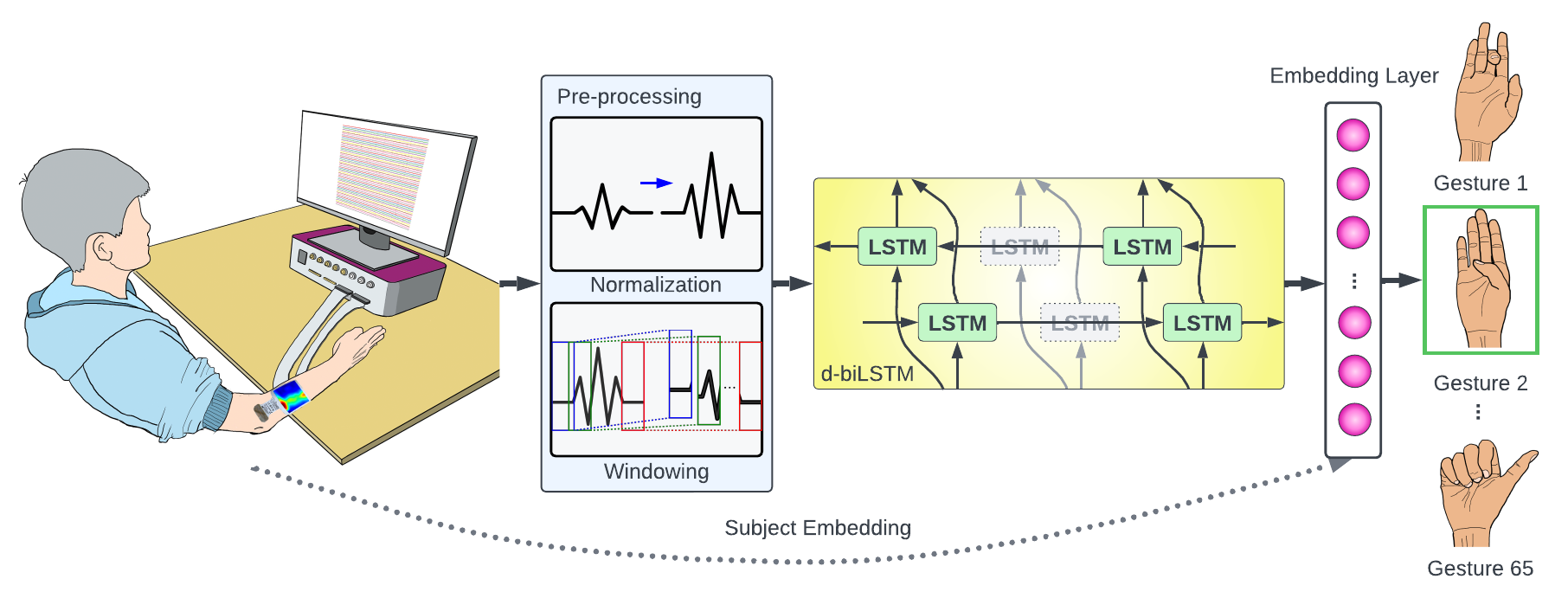}%
\end{wrapfigure}%
\begin{abstract}
Hand gesture recognition (HGR) has gained significant attention due to the increasing use of AI-powered human-computer interfaces that can interpret the deep spatiotemporal dynamics of biosignals from the peripheral nervous system, such as surface electromyography (sEMG). These interfaces have a range of applications, including the control of extended reality, agile prosthetics, and exoskeletons. However, the natural variability of sEMG among individuals has led  researchers to focus on subject-specific solutions. Deep learning methods, which often have complex structures, are particularly data-hungry and can be time-consuming to train, making them less practical for subject-specific applications. In this paper, we propose and develop a generalizable,  sequential decoder of transient high-density sEMG (HD-sEMG) that achieves 73\% average accuracy on 65 gestures for partially-observed subjects through subject-embedded transfer learning, leveraging pre-knowledge of HGR acquired during pre-training. The use of  transient  HD-sEMG before gesture stabilization allows us to predict gestures with the ultimate goal of counterbalancing system control delays. The results show that the proposed generalized models significantly outperform subject-specific approaches, especially when the training data is limited and there is a significant number of gesture classes. By building on pre-knowledge and incorporating a multiplicative subject-embedded structure, our method comparatively achieves more than 13\% average accuracy across partially-observed subjects with minimal data availability. This work highlights the potential of HD-sEMG and demonstrates the benefits of modeling common patterns across users to reduce the need for large amounts of data for new users, enhancing practicality.
\end{abstract}

\end{minipage}}}

\maketitle
\section{Introduction}
\label{sec:intro}
The increasing use of internet of things (IoT)  and investment in commercial augmented and virtual reality (AR/VR) applications suggest a growing demand for human-computer interfaces (HCI) \cite{noauthor_undated-vf,noauthor_ar/vr}. This demand is further highlighted by the growing population of people with disability and amputees in the United States \cite{Ziegler-Graham2008-yn} that underlines the importance of neurorobotic systems (e.g., exoskeletons and prosthetics) equipped with HCI.

Surface electromyography (sEMG) has been commonly used to register the activation of the peripheral nervous system and as part of non-invasive neural interfaces \cite{Merletti2016-wp, Raez2006-lm, cote-transfer,feature-alignment, Ahsan2009EMGSC, robust-pr,Phinyomark2013EMGFE, Sun2022-kf,ieeesensors-semg}. 
High-density surface EMG (HD-sEMG) is a variant of non-invasive sEMG, collected through arrays of densely-located  electrodes to provide a more detailed scan of the propagation of neural drive over space and a high-resolution representation of the muscle activity \cite{Sun2022-kf, hd-emg, Bahador2020-gk,Boschmann2012,ieeesensors-channels}. The aforementioned modalities have a wide range of applications, including gesture classification and tracking in HCI  \cite{cote-transfer,feature-alignment, robust-pr,Ahsan2009EMGSC,Phinyomark2013EMGFE,ieeesensors-hci}, and beyond, such as  the assessment of muscle function \cite{Bi2019-ht}, the diagnosis of neuromuscular disorders \cite{Raez2006-lm}, the evaluation of muscle fatigue \cite{Yousif_2019,Merletti2016-wp}. Machine Learning and Deep Learning (DL) have enabled the development of robust decoders for sEMG and HD-sEMG to detect the intended motor commands of users in various contexts. As described in the ``Prior Works'' section below, recent advances in DL have obtained remarkably accurate decoding of gestures from sEMG signals \cite{jafarzadeh_hussey_tadesse_2019,Xiong-review,dl-review,Ameri2018-bb, Ameri2019-ub, Asif2020-vw,real-emb-dl,3d-cnn,Chen2020-bp,d-CNN,Betthauser2019,Betthauser2020,Sun2022-kf, homo-dLSTM,hybrid,ViT-HGR,TFS-HGR,FS-HGR,Few-Shot}.

Despite the progress, a remaining key challenge of using DL for decoding sEMG is that sEMG signals vary significantly among various subjects \cite{robust-pr,ieeesensors-dist}. Thus,  individual subject characteristics such as age, muscle composition, skinfold thickness, and gesture styles and habits can all influence the mapping between the sEMG space and intended motions \cite{Ahsan2009EMGSC,robust-pr,feature-alignment,cote-transfer}.
This variability has made it unlikely for a single model to accurately predict a high number of gestures across multiple subjects without retraining and fine-tuning \cite{feature-alignment}.
For example, \cite{jafarzadeh_hussey_tadesse_2019} showed that a Convolutional Neural Network (CNN) used to decode 15 hand gestures achieved a high validation accuracy of 91.26\% in a subject-specific experiment but a low test accuracy of 48.40\% when tested on unseen subjects.

The poor performance on unseen subjects has necessitated the development of  subject-specific models.
As a result, a complex DL model needs to be fully retrained for each new subject using sizable labeled data from that individual. This exhaustive training and tedious data collection process raise questions about such DL techniques in terms of effectiveness and translation beyond research labs\cite{robust-pr}. The aforementioned challenge has motivated the research on potential generalizable approaches to decode sEMG.  
Early attempts at generalization (see \cite{Matsubara-bilinear,feature-alignment,transferHGR,cote-transfer,CAPSnet,robust-pr,FS-HGR,Few-Shot} described in the Prior Works section), mostly using multichannel bipolar sEMG, have demonstrated successes only on tasks with limited number of gestures (in the range of five to 18 gestures) and limited gesture complexity (focusing on gestures with highly distinguishable patterns).
In contrast, this work will attempt generalization for decoding 65 gestures, including complex and similar motions.
We will investigate the power of HD-sEMG with 128 channels that can capture muscle activity with high spatiotemporal resolution. This is done with the goal of detecting underlying patterns of muscle activation propagated in the space and time that can possibly be used for generalization.

\textbf{Contributions of this work:}  
The goal of this study is to push the boundaries of generalizable and agile gesture decoding by attempting to achieve high classification accuracy for a large number of gestures across different subjects via minimal HD-sEMG data while focusing on the transient phase of gesture conduction (to reduce decoding latency in the resulting HCI). We propose a dilated bi-directional long short-term memory (d-biLSTM) model that combines the advantages of temporal dilation and a bi-directional structure.
At root, our model is designed with the goal of overcoming the decoding complexity and inherent variability in sEMG signals among subjects.
Addressing this problem can significantly impact  clinical and practical applications of sEMG in HCI. For this purpose, our training approach (i.e., \emph{subject-embedded transfer learning for gesture prediction using transient HD-sEMG}) is composed of two phases: (a) training the base model to capture the common neurophysiological patterns of gesture performance from a limited number of subjects through ``common parameters" and (b) retraining the model on a new subject with limited available HD-sEMG data (referred to as ``partially-observed subject" in the rest of the paper) to find the subject-specific projection through common and ``subject-specific'' parameters.
In this work, the mapping of each subject index to the subject-specific projection is considered an embedding. In the first phase, the base model is trained with data from few subjects (referred to as pre-training subjects). Both common and subject-specific parameters are trained from scratch in this phase. During the retraining phase on a new subject, the common parameters learned in the first phase are used as the initial condition, and an embedding vector for the new subject is initialized by the average of embedding vectors corresponding to pre-training subjects.
Unlike traditional transfer learning (TL), the proposed method enables subject-specificity in the pre-training set by incorporating multiplicative subject embedding.
We demonstrate that the proposed method has several significant advantages over pure subject-specific models and previous traditional TL:  
\begin{itemize}
\item \emph{Generalization with HD-sEMG signals and large numbers of gestures}.
 As described in the Prior Works section below, earlier efforts on generalizing to new subjects have mostly demonstrated success with limited number of gestures.  In contrast, this paper presents  generalization performance on 65 gestures. This is motivated as we use HD-sEMG which introduces higher information rate for finding common patterns. Most of the prior works (with the exception of \cite{transferHGR}) attempted generalization only using sparsely located bipolar sEMG rather than HD-sEMG.
\item \emph{Generalization with transient-phase HD-sEMG}. The prior work on the topic of generalization uses the plateau phase of sEMG during gesture conduction when the signal is mostly stable. This, in general, can introduce extra latency in the HCI system and inaccuracy during transition from one gesture to another. In this paper, for the first time we approach the more challenging problem of generalization on transient phase of HD-sEMG with the goal of predicting the upcoming gestures and reducing the latency in HCI. 
\item \emph{Generalization with minimal new data:}  
The proposed model reaches an average accuracy of 73\% across partially-observed subjects when having access to a limited number of repetitions per gesture during the retraining phase. More specifically, in this study, the challenging problem of \textit{single-repetition} decoding has been addressed, requiring retraining our model for each subject using one repetition of data. This means that we can achieve 13\% more accuracy compared to the state-of-the-art subject-specific counterpart while having access to only 25\% of data.
\item \emph{Lightweight bi-directional LSTM:} In addition to the multiplicative embedding, the proposed model is compact with ~79K trainable parameters (e.g., compared to 1.7 million parameters of CapsNet\cite{CAPSnet}). It should be noted that for cloud computing, the compactness of the model structure saves substantial computational resources for rapid upgrades of the models on the cloud and individual devices. Moreover, model compactness enhances the practicality in terms of the implementation of portable hardware, such as HCI controllers.
\end{itemize}

The results of this paper support the hypothesis that subject-embedded TL can indeed improve the HGR accuracy on new subjects with limited calibration. We have observed that the proposed generalized model consistently and significantly outperforms both purely subject-specific models as well as traditional TL-based models for all rates of data availability. The accuracy improvement gap is particularly large when sufficient training data (i.e., sufficient repetitions for each gesture) from the partially-observed subject is not available.

The remainder of this paper is organized as follows: Section \ref{prior works} reviews the prior studies considering the problem of HGR from sEMG signals, including subject-specific models and attempts towards generalization to new subjects. In section \ref{sec:embedded}, our proposed subject-embedded transfer learning strategy and the intuition it relies on are explained. Section \ref{sec:db-dp} introduces the dataset and pre-processing scheme used in this study. Section \ref{sec:model} provides an overview of the proposed model architecture. In section \ref{sec:experiments-results}, we describe the training and retraining configurations, the experiments conducted, and the corresponding results. Section \ref{sec:comparative} is dedicated to comparing our proposed model to other LSTM-based architectures and analyzing the benefits of the proposed embedding-based generalization approach. We also evaluate our method on the steady-state (i.e., plateau) phase of HD-sEMG signals. Finally, in section \ref{sec:conclusions} we summarize our observations and draw conclusions.

\section{Prior Works: From Subject-Specific to Generalization}\label{prior works}

\textit{ \bf Subject-specific Models:}
Subject-specific models for sEMG are trained for each individual and require complete retraining before being used on a new subject. Previous subject-specific studies have primarily used feature extraction methods and traditional machine learning techniques. These are just a few examples of the classic efforts in this field \cite{Simao2019-kb}. Linear discriminant analysis (LDA) \cite{Phinyomark2013EMGFE,Phinyomark2012FeatureRA}, Gaussian Naive Bayes \cite{byfield-realtime}, clustering-based algorithms \cite{momen-emg-userselected}, decision trees, and hidden Markov models \cite{Zhang-accelerometer} have been used to decode hand gestures from sEMG signals with high accuracies.
sEMG signals are complex and variable, with a non-stationary and non-linear relationship to muscle contractions \cite{nature-database}. These characteristics make it difficult to model the relationship between the signal and the gesture explicitly, especially for a large number of gestures. As a result, DL algorithms have been increasingly used to decode sEMG signals into gesture classes \cite{Xiong-review,dl-review}.  In the following, we provide some examples of the use of DL for this purpose.

CNNs \cite{Ameri2018-bb, Ameri2019-ub, Asif2020-vw, real-emb-dl} and CNN-inspired architectures such as temporal convolutional networks (TCNs) \cite{Betthauser2019,Betthauser2020}, compact CNNs (EMGNet) \cite{Chen2020-bp}, 3D CNNs \cite{3d-cnn}, and dilated CNNs \cite{d-CNN} have been studied extensively towards myoelectric control and pattern recognition in the past few years. These models have reached high accuracies (e.g., 97\%) depending on the number of gestures- during subject-specific studies. Long short-term memory (LSTM) networks have demonstrated satisfactory performances as well. Such architectures have been significantly improved by introducing temporal dilation \cite{Sun2022-kf, homo-dLSTM}, reaching an accuracy of 83\% for decoding 65 gestures from HD-sEMG signals. Hybrid architectures, combining CNNs and LSTMs are shown to be accurate as well \cite{hybrid}.
DL algorithms require a large amount of labeled data for training in order to achieve satisfactory accuracy, and obtaining such data can be impractical in many cases \cite{cote-transfer}.
Transformer-based and Few Shot Learning (FSL)-based frameworks have been proposed to address the elongated training time and the limited data availability problems of DL models, respectively \cite{ViT-HGR,TFS-HGR,FS-HGR,Few-Shot}.

\textbf{Efforts at Generalization and Their Limitations:}
Several works have attempted to develop generalized models, but remain limited in various aspects.
Authors of \cite{Matsubara-bilinear} propose a bilinear model that can detect five gestures with an accuracy of 73\% by an adaptation process.
In addition, in \cite{feature-alignment},  an Unsupervised Domain Adaptation (UDA) is proposed to classify six gestures with an average accuracy of 90.41\%. 
TL can be used to improve model performance in a target domain through knowledge from the source domain. Authors of \cite{transferHGR} propose a TL strategy with majority voting that reaches an average accuracy of 95.97\% for 12 basic finger movements in CapgMyo-DBc (a HD-sEMG database \cite{Du_Sensors_2017}). These three studies \cite{Matsubara-bilinear,feature-alignment,transferHGR} only consider a few number of gestures. 
The effect of TL on improving the accuracy of a convolutional network architecture to detect 18 gestures for new subjects is studied in \cite{cote-transfer}. They report an accuracy of 68.98\% when given four repetitions of new data. Besides the limited number of gestures, more repetitions are required to calibrate this model for new subjects, in comparison to our proposed model. 
Authors of \cite{CAPSnet} propose the dilated efficient capsular neural network (CapsNet) that can predict 17 gestures from the transient phase of sEMG signals with an accuracy of 78.3\%. The disadvantages of this model include the large number of trainable parameters and low number of gestures.
The authors of \cite{robust-pr} report both subject-specific and generalized (inter-subject) accuracy for static and dynamic gestures.
The authors of \cite{robust-pr} have introduced a CNN model named the Multitask Dual-Stream Supervised Domain Adaptation Network (MDSDA) that exhibits long-term robustness and adaptability to multiple subjects,  reporting an inter-subject accuracy of 97.2\% in detecting ten gestures.
The low number of gestures and high complexity of the model are limitations of the study  in \cite{robust-pr}.
Authors of \cite{FS-HGR,Few-Shot} use FSL to improve accuracy on new subjects. They report accuracies in the range of 76.39\%-81.29\% based on different architectures they use for five-way five-shot experiments. Few number of gestures (five-way) and requirement for more repetitions (five-shot) of new data are the limitations of these studies.
In a very recent study \cite{uda2023},  domain generalization and UDA were integrated into a single framework that detects seven gestures from HD-sEMG signals. Similarly, the low number of gestures is the limitation of this study besides the low-complexity of the targeted gestures.
\section{Subject-Embedded Transfer Learning}
\label{sec:embedded}
We briefly describe the general principle of the proposed subject-embedded TL and how it contrasts to other methods.
Consider a general  problem of predicting some target $y$ from an input $x$.
In the sEMG problem, $y$ will be the gesture index and $x$ will be an array representing the multi-channel data collected over some time interval. Let $u$ denote a subject index. One simple predictor would be of the form:
\begin{equation} \label{eq:fsing}
    \widehat{y} = f(x,\theta),
\end{equation}
where $\widehat{y}$ is the prediction of the target $y$, and $f(x,\theta)$ is a function with parameters $\theta$. For example, $f(x,\theta)$ could be a neural network with input $x$ and $\theta$ would be weights and biases.
By a \textbf{common model}, we mean that we learn a single common parameter $\theta$ for all subjects $u$. The obvious drawback with a common model is that it cannot capture subject-specific characteristics of the mapping.  The other extreme would be a \textbf{subject-specific model} where one set of parameters $\theta$ is learned for each subject $u$. 
As mentioned in the Introduction section, the challenge of subject-specific models is that they require significant training data for each subject.

One approach to reduce the data required for subject-specific models is to use what we will call \textbf{standard TL}. In this method, one typically first selects one or more \emph{pre-training subjects} and learns a \emph{common base model} $\widehat{y} = f(x,\theta^0)$ for these pre-training subjects where $\theta^0$ represents the \emph{base parameters}.  Then, given a new subject $u$, the parameters $\theta^0$ are finely-tuned to obtain a new subject-specific parameter $\theta(u)$.  The simplest method is to divide the parameters into components $\theta = (\theta_1,\theta_2)$.  For example, $\theta_1$ are the weight and biases for the initial layers, and $\theta_2$ are the parameters for the final (generally fully-connected or FC) layers.  In the pre-training phase, we learn base parameters $\theta^0=(\theta_1^0, \theta_2^0)$. For the new subject, we freeze $\theta_1^0$ and only learn a subject-specific component, $\theta_2(u)$, thereby reducing the parameters to be learned. The problem in this method, is that the base model is not subject-specific, and therefore, may not be able to provide a good fit over a large pre-training set.

 For the proposed subject-embedded TL, we similarly divide the parameters into two components, $\theta = (\theta_1,\theta_2(u))$. In the pre-training phase, we learn a parameter $\theta^0=(\theta_1^0,\theta_2(u))$, where the first component, $\theta_1^0$, is common to all pre-training subjects.  However, unlike standard TL, the second parameters, $\theta_2(u)$, is dependent on the subject index $u$ within the pre-training set. The mapping of the subject index $u$ to the parameters $\theta_2(u)$ can thus be seen as an embedding of the subject in some parameter space.  This embedding enables the base model to have a subject-specific component.

For a new subject, $u'$, not in the pre-training set, we run traditional gradient-descent learning on the data from a new subject where:
(1) we initialize the first component $\theta_1 = \theta_1^0$, the common parameters in the base model; and
(2) we initialize the second component, $\theta_2(u')$ to the average of $\theta_2(u)$ for $u$ in the pre-training set. The initialization of $\theta_1=\theta_1^0$ implicitly captures the common aspects of the model from the pre-training set, while the search over $\theta_2(u')$ helps capture the subject-specific characteristics of the new subject.
\section{Database}
\label{sec:db-dp}
\subsection{Data Acquisition}
\label{subsec:db}
\begin{figure}[htb]
    \centering
    \subfloat[Little and ring fingers bend.]{%
		\includegraphics[width=0.7\columnwidth]{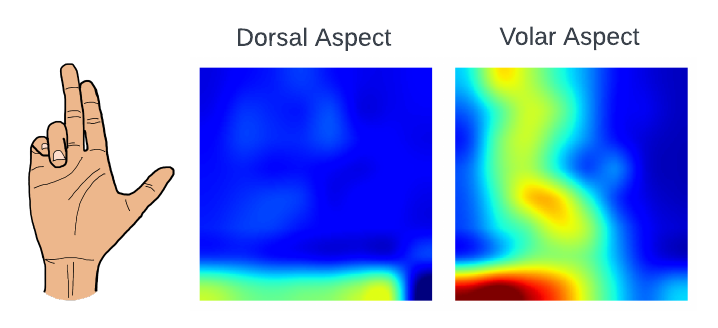}}\\
	\subfloat[All fingers extension (without thumb).]{%
		\includegraphics[width=0.7\columnwidth]{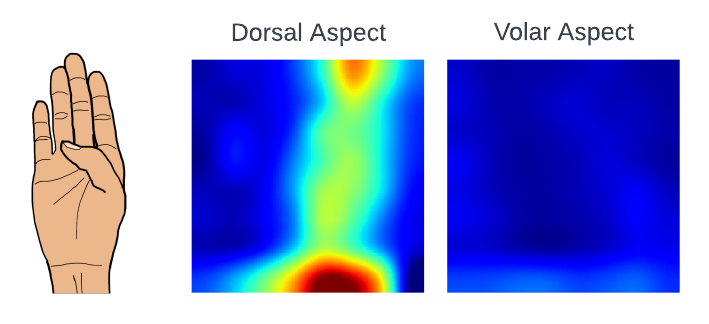}}
    \caption{Two example gestures with corresponding heatmaps that are the root mean square of a 200ms window.}
	\label{fig:example-gestures}
	\vspace{-1em}
\end{figure}

\begin{figure}[htb]
    \centering
    \includegraphics[width=0.7\columnwidth]{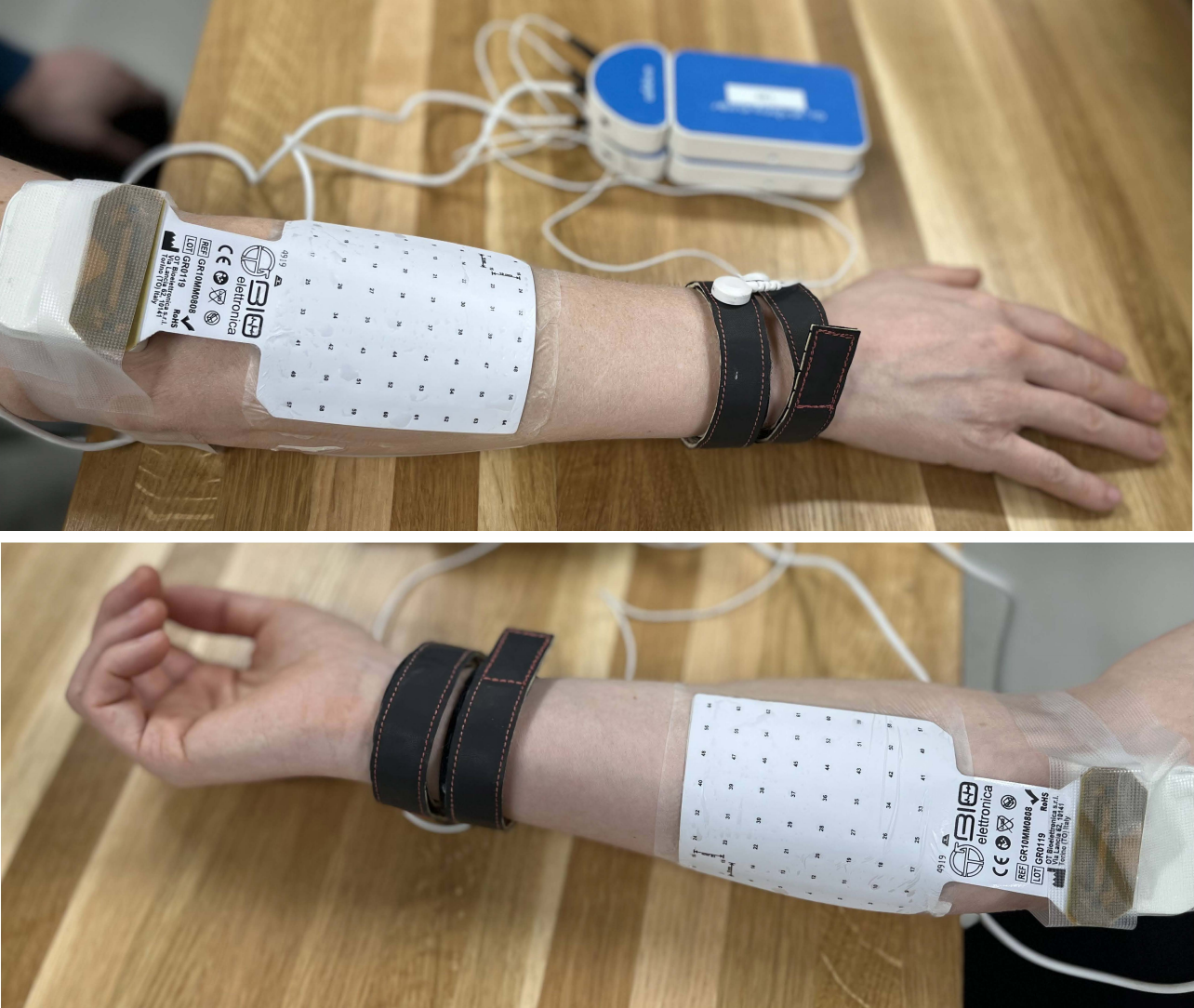}
    \caption{The placement of two 8 $\times$ 8 electrode grids, with one on the dorsal aspect (outer forearm) and the other on the volar (inner forearm) aspect of the forearm \cite{Sun2022.02.25.481922}.} 
    \label{fig:placement}
    \vspace{-1em}
\end{figure}

Our work aims to develop a robust, multi-functional HCI control system capable of supporting a diverse range of control tasks through HD-sEMG data. To this end, the study uses a publicly available open-source HD-sEMG database containing 65 isometric hand gestures \cite{nature-database}. HD-sEMG data provide rich spatiotemporal information about underlying muscle activity and are particularly useful in recognizing a large number of gestures. The database includes 16 gestures with one degree of freedom (DoF), 41 gestures with two DoFs, and eight gestures with three DoFs, encompassing a range of finger and wrist movements such as bending, stretching, rotating, grasping, pointing, and pinching. Fig. \ref{fig:example-gestures} shows two example gestures with their corresponding muscle-activity heatmaps. The signals were collected by a Quattrocento (OT Bioelettronica, Torino, Italia) biomedical amplifier through two 8 $\times$ 8 electrode grids (128 sensors in total) positioned on the volar and dorsal aspects of the forearm at a sampling rate of 2048 Hz (see Fig. \ref{fig:placement} for visualization of electrode placement \cite{Sun2022.02.25.481922}). This database was collected from 20 able-bodied subjects (14 men and six women with average age of 35) who were instructed to perform each gesture for five repetitions, each lasting five seconds, with a five-second inter-repetition rest period. The plateau phase of the repetitions is often used in gesture recognition due to the stability of muscle contraction during gesture maintenance, introducing control delay in practical applications. However, this study focuses only on transient-phase signals, which include the most dynamic muscle activity, to design an agile HCI control system that can begin recognizing gestures as soon as a user initiates one. HGR on transient data transforms gesture detection into gesture prediction, minimizing control delay.

\subsection{Data Pre-processing}
\label{subsec:preprocess}
The raw HD-sEMG signals from the two 8 $\times$ 8 electrode grids are flattened and concatenated to form 128-channel signals suitable to our proposed sequential model. The total data can then be represented  as a tensor, $x[n,i,t]$, where $n$ is the sample index, $i$ is the channel and $t$ is the time index within the sample.  Each sample is the data from one repetition, so we will use the term repetition and sample interchangeably. Magnitudes of muscle signals such as sEMG vary according to muscle type, length, and velocity \cite{Roberts2008-iv}. sEMG can benefit from normalization in the sense that the signals collected from each electrode can contribute equally during model training \cite{Singh2020-xr}. We normalize the raw signals using z-score transformation via the means and standard deviations from only the training data. The standardization method converts the raw signals to a common scale such that the standardized signals have a zero mean and unit standard deviation. We then 
obtain scaled data:
\begin{equation}
    v[n,i,t] = \frac{x[n,i,t] - \mu[i]}{\sigma[i]}.
\end{equation}
where
\begin{subequations}
    \begin{align}
        \mu[i] &:= \frac{\sum_{n \in N_{\rm tr}, t} x[n,i,t]}{N},\\
    \sigma[i] &:= \sqrt{\frac{\sum_{n \in N_{\rm tr}, t}(x[n,i,t] - \mu[i])^2}{N}}.
    \end{align}
\end{subequations}

Following the previous work \cite{Sun2022-kf}, we define the duration of the transient phase as the first 0.5 seconds of each repetition, according to the average force signals of each gesture. Windowing is a commonly-used data augmentation technique for model performance and generalization enhancement in sEMG-based gesture detection. A repetition of the normalized signals is segmented into multiple overlapping windows before being fed into the proposed model. We use a window size of 200 ms and a stride of 10 ms to meet the requirement of real-time control \cite{Hassan2020-nh, Shi2018-vk, Aranceta-Garza2019-uz, Wahid2020-ne}. The data pre-processing can be visualized as the upper box in Fig. \ref{fig:process}.

In the pre-training phase, the train-test division is determined by assigning repetitions 1, 3, and 5 to the training set, and the remaining repetitions (2 and 4) to the testing set. To evaluate the capability of the generalized models on partially-observed subjects given different data availability in retraining, we calibrate the generalized models on any selection of one (33\%), two (67\%), and all three (100\%) of the training repetitions. The train-test split for subject-specific model training follows the setups in the retraining phase.
\section{Model Architecture}
\label{sec:model}
We propose the d-biLSTM model that combines the advantages of temporal dilation and a bi-directional structure. The model consists of three components: a three-layer d-biLSTM, a classifier with FC layers and dropout, and an embedding layer that captures subject dependencies in the generalized model. We will now provide a brief overview of these components.

\subsubsection{biLSTM}
The exploding and vanishing gradient issue in recurrent neural networks (RNNs) has been well-studied in the literature and is often addressed through the use of LSTM cells \cite{Hochreiter1997LongSM,Quivira-RNN}. LSTM introduces additional gating mechanisms that enable the model to selectively retain or forget information, allowing it to better capture long-range dependencies in the input data. Inspired by the work in \cite{homo-dLSTM}, we introduce temporal dilation in the LSTM architecture to further improve its ability to capture long-term dependencies in the input data. Temporal dilation allows for an expansion of the temporal receptive field (in the time series) without increasing the number of parameters and indeed, reducing the computational cost, making it an effective method for capturing longer-range dependencies and complex input sequences. In addition, in order to fully utilize all of the past and future information available within a specific signal window, we utilize a bi-directional LSTM (biLSTM) structure instead of a standard LSTM. A biLSTM processes the input sequence in both the forward and backward directions, allowing it to integrate contextual information from both past and future time steps within the processing window. Our experiments demonstrate that the biLSTM can achieve comparable accuracy to the LSTM while requiring fewer trainable parameters. The model consists of three d-biLSTM layers, each containing 32 hidden units and dilated with a factor of three, such that the next d-biLSTM layer has $\frac{1}{8}$ connected LSTM cells compared to the current layer (details of homogeneous temporal dilation can be referred to \cite{homo-dLSTM}). The combination of temporal dilation and bi-directional processing enables the d-biLSTM model to effectively learn and classify the complex sequential sEMG data. Each d-biLSTM layer has a set of forward and backward outputs which are added before being passed as the input to the next layer. The final forward and backward hidden states are concatenated (yielding a 64 dimensional vector) before being fed into the classifier module.
\subsubsection{Embedding}

The weights of an embedding layer create a matrix that serves as an encoder of subject-specific information, resembling a lookup mechanism. The dimension of the embedding matrix can be adjusted based on the number of subjects and the specific model architecture in which it will be used. When given a subject index, a row from the embedding matrix corresponding to that specific subject is extracted and used in the model. In this study, we use a multiplicative embedding structure, where the extracted row is multiplied with the output of the first FC layer in the classifier module. The embedding rows have a dimension of 32 to match the structure of the classifier. This allows the model to effectively capture subject dependencies in the input data and improve performance on the classification task.

\subsubsection{Classifier}
As part of the classifier, first, a FC layer with the hyperbolic tangent activation function is used to decrease the output dimension of the d-biLSTM module from 64 to 32. It is followed by a dropout layer with a rate of 0.2 to avoid over-fitting the training data. The resulting vector is multiplied by the embedded vector extracted from the embedding module, given the subject index.
A final FC layer with the Softmax activation function is used to assign probabilities to various gesture classes. The predicted label corresponds to the gesture with the highest output probability.

\begin{figure*}
    \centering
    \includegraphics[width=0.95\linewidth]{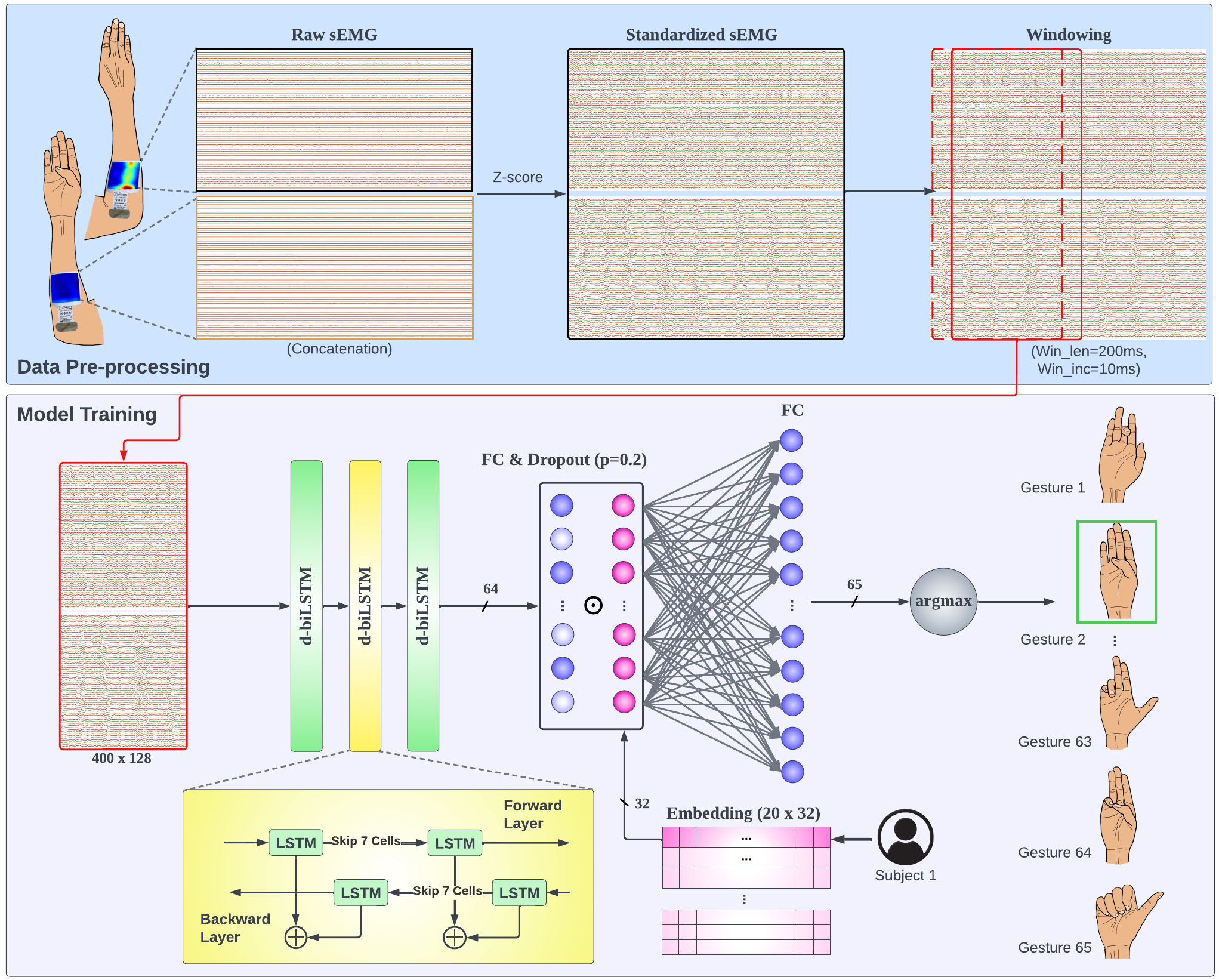}
  \caption{The figure illustrates the proposed pipeline for gesture prediction using HD-sEMG data. The blue area represents the data pre-processing steps, including sensor flattening and windowing. The purple area shows the proposed d-biLSTM model, which takes each window of normalized HD-sEMG data as input and predicts the 65 gestures according to the maximum predicted probability output from the softmax function. The generalized model (shown in the figure) includes a subject embedding layer (shown in magenta), which captures subject-specific dependencies in the data, while the subject-specific counterpart excludes this layer. This allows the model to effectively learn and classify sequential data for a wide range of subjects.}\vspace{-0.3cm}
    \label{fig:process}
\end{figure*}

In this study, two-phase generalization is conducted on the model with the embedding layer (named generalized model), whereas for the conventional subject-specific model that is only trained in one phase, no embedding layer is considered. Fig. \ref{fig:process} demonstrates the data pre-processing and model training pipelines. The lower box (i.e., Model Training) shows the proposed model structure.

\section{Experiments and Results}
\label{sec:experiments-results}
\label{subsec:experiments}
 In this section, experimental results are reported regarding various phases of training for the proposed generalized models in comparison with the performance of the corresponding subject-specific models. Also, additional results will be reported regarding the effect of data reduction and task complexity (i.e. number of gestures) on the performance. 
It will be shown that  (a)  the proposed generalized model outperforms its subject-specific counterpart more significantly when predicting a higher number of gestures with fewer available data, benefiting from the proposed TL which includes the multiplicative embedding layer through weight initialization in retraining; (b) The pre-trained weights represent the HGR pre-knowledge captured from known subjects, resulting in faster convergence for a new subject (100 epochs compared to 200) and reducing the chances of ending up at local minima.  

In all experiments, Adam optimizer with learning rate $1e-3$, $\beta_1=0.9$, $\beta_2=0.999$, and $\epsilon=1e-08$ is used. The proposed generalized model is pre-trained for 200 epochs with patience 40 on a number of subjects (referred to as the pre-training subjects). It is later retrained for only 100 epochs on a new subject (i.e., partially-observed subject). Moreover, the embedding vector of the partially-observed subject is initialized as the average of vector values for the pre-training subjects before the retraining phase. Categorical cross-entropy is used as the loss function and validation categorical accuracy is monitored. The number of training parameters is about 79,000 for the generalized and subject-specific models, depending on the number of gestures. Table \ref{table:comparison} summarizes the number of parameters and training epochs for two phases of the generalized model and one phase of subject-specific model training when detecting 65 gestures. The subject-specific model in each gesture detection task is trained from scratch for 200 epochs.  
\begin{table}[htb]
    \renewcommand{\arraystretch}{1.3}
    \caption{Comparison of the number of trainable parameters and training epochs when predicting 65 gestures.}
    \label{table:comparison}
    \centering
    \begin{tabular}{c||c|c|c }
        \hline
        \bfseries   & \bfseries \shortstack{Subj. training} & \bfseries \shortstack{Gen. pre-training} & \bfseries \shortstack{Gen. retraining} \\
        \hline\hline
        Parameters & 78,721 & 78,881 & 78,753\\
        Epochs & 200 & 200 & 100 \\
        \hline
    \end{tabular}
    \vspace{-1em}
\end{table}

\subsection{Pre-training}
\begin{figure*}[htb]
    \centering
    \includegraphics[width=0.8\linewidth]{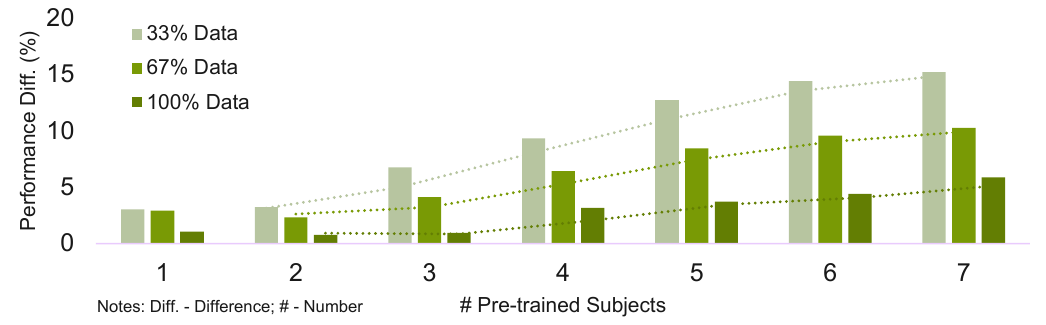}\\ 
     \caption{Performance difference when given a varying number of pre-training subjects. The performance difference (y-axis) is the average accuracy difference between a generalized model and its subject-specific counterpart across partially-observed subjects based on the same percentage of retraining data. The lighter the color shade, the less the retraining data. The dotted lines indicate the moving averages as the trend of the ascending performance gap as more known subjects are included in pre-training.}
    \label{fig:pretrain-subs}
\end{figure*}

As an initial step, the proposed generalized model needs to be pre-trained on a number of subjects. At this phase, we have access to sufficient recordings of gestures from multiple subjects.
To determine the optimal number of subjects to use in the pre-training phase of the generalized model, we conduct a series of experiments in which the model is pre-trained on one to seven subjects for the task of detecting 65 gestures. We select a diverse group of subjects in terms of subject-specific accuracy to mimic the real-world scenario in which a variety of subjects might be chosen for pre-training.
Table \ref{table:pretraing-subjects} shows the generalized average accuracy of detecting 65 gestures evaluated on the remaining 13 subjects on three levels of data availability (33\%, 67\%, and 100\% which correspond to one, two, and three retraining repetitions per gesture, respectively). It can be seen that increasing the number of pre-training subjects improves the generalized accuracy, and thus, the generalized model consistently outperforms the subject-specific model, especially when insufficient data is available for a new subject.
Fig. \ref{fig:pretrain-subs} shows the accuracy improvement of the generalized model over the subject-specific model with respect to the number of subjects used in the pre-training phase of the generalized model. It can be observed that the accuracy improvement from adding the sixth and seventh subjects is less significant. Given the limited number of subjects available in the dataset (20 subjects), we use a maximum of five subjects to pre-train the generalized model in all other experiments in order to have sufficient test subjects for statistical analysis. 
\begin{table*}[htb]
    \renewcommand{\arraystretch}{1.3}
    \caption{Comparison of average accuracy for partially-observed subjects (evaluated on 13 subjects) for predicting 65 gestures.}
    \label{table:pretraing-subjects}
    \centering
    \begin{tabular}{l|c|c|c|c|c|c|c|c}
    \hline
    \multicolumn{9}{c}{Average accuracy per selection of pre-training subjects when detecting 65 gestures} \\
    \hline
    \multicolumn{1}{c||}{\bfseries Data access} & \multicolumn{7}{c||}{\bfseries Number of pre-training subjects for generalized model} & \multicolumn{1}{c}{\bfseries Subject}\\
    \cline{2-8}
    \multicolumn{1}{c||}{(\%)} & \multicolumn{1}{c||}{\bfseries 1} & \multicolumn{1}{c||}{\bfseries 2} & \multicolumn{1}{c||}{\bfseries 3} & \multicolumn{1}{c||}{\bfseries 4} & \multicolumn{1}{c||}{\bfseries 5} & \multicolumn{1}{c||}{\bfseries 6} & \multicolumn{1}{c||}{\bfseries 7} & \multicolumn{1}{c}{\bfseries specific}\\
    \cline{1-9}
    \multicolumn{1}{c||}{33} & \multicolumn{1}{c||}{43.28}  & \multicolumn{1}{c||}{43.46} &  \multicolumn{1}{c||}{47.00}& \multicolumn{1}{c||}{49.58}&  \multicolumn{1}{c||}{52.95}&  \multicolumn{1}{c||}{54.67} & \multicolumn{1}{c||}{55.47} & \multicolumn{1}{c}{40.20} \\
    \multicolumn{1}{c||}{67}  &  \multicolumn{1}{c||}{61.49} &  \multicolumn{1}{c||}{60.86} &  \multicolumn{1}{c||}{62.67} &  \multicolumn{1}{c||}{64.99} &  \multicolumn{1}{c||}{67.03} &  \multicolumn{1}{c||}{68.13} &  \multicolumn{1}{c||}{68.83} & \multicolumn{1}{c}{58.53}\\
    \multicolumn{1}{c||}{100}  &  \multicolumn{1}{c||}{71.16} &  \multicolumn{1}{c||}{70.87} &  \multicolumn{1}{c||}{71.04} & \multicolumn{1}{c||}{73.26} &  \multicolumn{1}{c||}{73.85} &  \multicolumn{1}{c||}{74.53} & \multicolumn{1}{c||}{75.96} & \multicolumn{1}{c}{70.08}\\
    
    \hline
    \end{tabular}
\end{table*}

\subsection{Data reduction experiments}
\begin{figure}[htb]
    \centering 
         \includegraphics[width=0.8\linewidth]{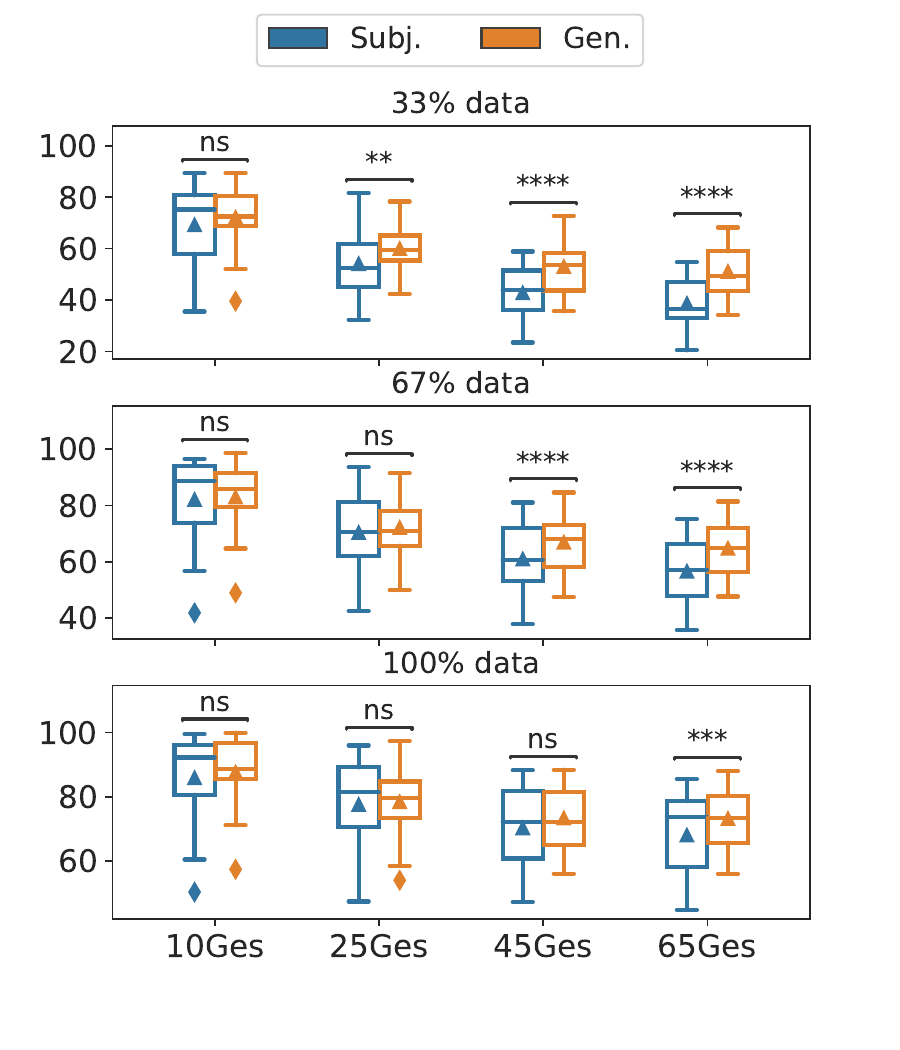}\\
\vspace{-0.5cm}\caption{Comparison of models during data reduction experiments when the generalized model is pre-trained on five random subjects. The compared distribution pairs are dependent, as they are the accuracies of the same subjects derived from the generalized and subject-specific models. Also, the accuracy differences are not normally distributed. Thus, the Wilcoxon signed-rank test with $\alpha = 0.05$ is employed as the statistical hypothesis test. X and Y axes denote the number of gestures and accuracy, respectively. Each box plot includes 15 data points that represent 15 partially-observed subjects.}
    \label{fig:comparison-task}
\end{figure}

As mentioned earlier, a main challenge in developing PR models is the requirement of a large dataset to train complex models for each new subject. In this section, we analyze the effect of available training data on the final accuracy. 
This reduction is based on the available training repetitions and will be analyzed in the following setting:
\begin{itemize}
    \item The generalized model is adequately pre-trained on five random subjects using all three repetitions \{1,3,5\}. 
    \item After proper initialization of the embedding vector corresponding to a new subject, the generalized model is retrained on a subset of \{1,3,5\} repetitions of that subject (i.e., the partially-observed subject). The subsets roughly measure to \%100 (all three repetitions), \%67 (two repetitions), and \%33 (one repetition) of the data.
    \item The subject-specific model is trained from scratch on the same subset as the retraining of the generalized model.
    \item Final accuracies of both models are evaluated on repetitions \{2,4\} of that subject.    
\end{itemize}
The results of these experiments for the task of predicting 65 gestures are demonstrated in the right-most column of Fig. \ref{fig:comparison-task}. Noting that retraining the proposed generalized model converges faster than training a subject-specific model from scratch (Table \ref{table:comparison}), we also observe that for a high number of gestures, without having access to sufficient samples from the new subject, the proposed model more significantly outperforms the subject-specific counterpart. In fact, given only one repetition of each of the 65 gestures for a new subject, our proposed model achieves a prediction accuracy that is about 13\% higher than the subject-specific accuracy.
\subsection{Comparing to traditional TL}
In order to emphasize the advantage of using a subject-embedded structure, we also implement a traditional TL scheme for detecting 65 gestures by pre-training the model excluding the embedding layer on five random subjects (subjects 1, 6, 10, 11, and 14), and then after freezing all layers except the FC layers of the classifier, we retrain the model on various amounts of data from a new subject. The training configurations match those of the generalized model for a fair comparison. The results are reflected in the last column of Table \ref{table:final-results}. The results show that even by 100\% retraining on a new subject, the traditional TL model has a lower accuracy (around 9\% less) compared to the generalized model that is only retrained with 33\% data.

\subsection{Task complexity experiments}
To investigate the effect of task complexity on model performance, we conduct experiments on the proposed generalized model on various numbers of gestures, ranging from 10 to 65. For each experiment, five random subjects (subjects 1, 6, 10, 11, and 14) are selected for the pre-training phase and the remaining subjects are used for testing.  
Table \ref{table:final-results} shows the final test accuracy for different amounts of retraining data. 
It is important to note that the testing subjects for Table \ref{table:final-results} are different from those in Table \ref{table:pretraing-subjects}, so the final average accuracy for 65 gestures is not directly comparable.
The results in Table \ref{table:final-results} and Fig. \ref{fig:comparison-task} show that the performance gap between the generalized and subject-specific models increases as the task complexity increases. The largest gap was observed when the models are asked to predict 65 gestures using only 33\% of data from a partially-observed subject. This suggests that the generalized model becomes more effective as the number of gestures to be detected increases. In this case, the subject-specific model would require more  data for accurate classification.

\begin{table*}[htb]
    \renewcommand{\arraystretch}{1.3}
    \caption{Comparison of average accuracy for partially-observed subjects (evaluated on 15 subjects). Pre-trained on five random subjects. Subj., gen., and TL refer to subject-specific, generalized, and traditional TL models, respectively.}
    \label{table:final-results}
    \centering
    \begin{tabular}{l|c|c|c|c|c|c|c|c|c}
    \hline
    \multicolumn{10}{c}{Average accuracy per gesture detection task} \\
    \hline
    \multicolumn{1}{c||}{\bfseries Data access} & \multicolumn{2}{c||}{\bfseries 10} & \multicolumn{2}{c||}{\bfseries 25} & \multicolumn{2}{c||}{\bfseries 45} & \multicolumn{3}{c}{\bfseries 65}\\
    \cline{2-10}
    \multicolumn{1}{c||}{(\%)} & \multicolumn{1}{c|}{\bfseries subj.} & \multicolumn{1}{c||}{\bfseries gen.} & \multicolumn{1}{c|}{\bfseries subj.} & \multicolumn{1}{c||}{\bfseries gen.} & \multicolumn{1}{c|}{\bfseries subj.} & \multicolumn{1}{c||}{\bfseries gen.} & \multicolumn{1}{c|}{\bfseries subj.} & \multicolumn{1}{c|}{\bfseries gen.} & \multicolumn{1}{c}{\bfseries TL} \\
    \hline
    \multicolumn{1}{c||}{33}  & \multicolumn{1}{c|}{69.20} & \multicolumn{1}{c||}{72.10} & \multicolumn{1}{c|}{54.07} & \multicolumn{1}{c||}{59.94} & \multicolumn{1}{c|}{42.72} & \multicolumn{1}{c||}{53.01} & \multicolumn{1}{c|}{38.56} & \multicolumn{1}{c|}{51.05} & \multicolumn{1}{c}{32.01} \\
    \multicolumn{1}{c||}{67}   & \multicolumn{1}{c|}{82.02} & \multicolumn{1}{c||}{82.94} & \multicolumn{1}{c|}{70.30} & \multicolumn{1}{c||}{72.11} & \multicolumn{1}{c|}{60.94} & \multicolumn{1}{c||}{66.87} & \multicolumn{1}{c|}{56.58} & \multicolumn{1}{c|}{64.70}  & \multicolumn{1}{c}{38.34} \\
    \multicolumn{1}{c||}{100}  & \multicolumn{1}{c|}{85.88} & \multicolumn{1}{c||}{87.61} & \multicolumn{1}{c|}{77.56} & \multicolumn{1}{c||}{78.42} & \multicolumn{1}{c|}{70.29} & \multicolumn{1}{c||}{73.36} & \multicolumn{1}{c|}{68.04} & \multicolumn{1}{c|}{73.22}  & \multicolumn{1}{c}{41.87}\\
    \hline
    \end{tabular}
\end{table*}
\section{Comparative Study}
\label{sec:comparative}
\begin{figure}[ht]
    \centering
    \subfloat[33\% Data]{%
		\includegraphics[width=0.48\columnwidth]{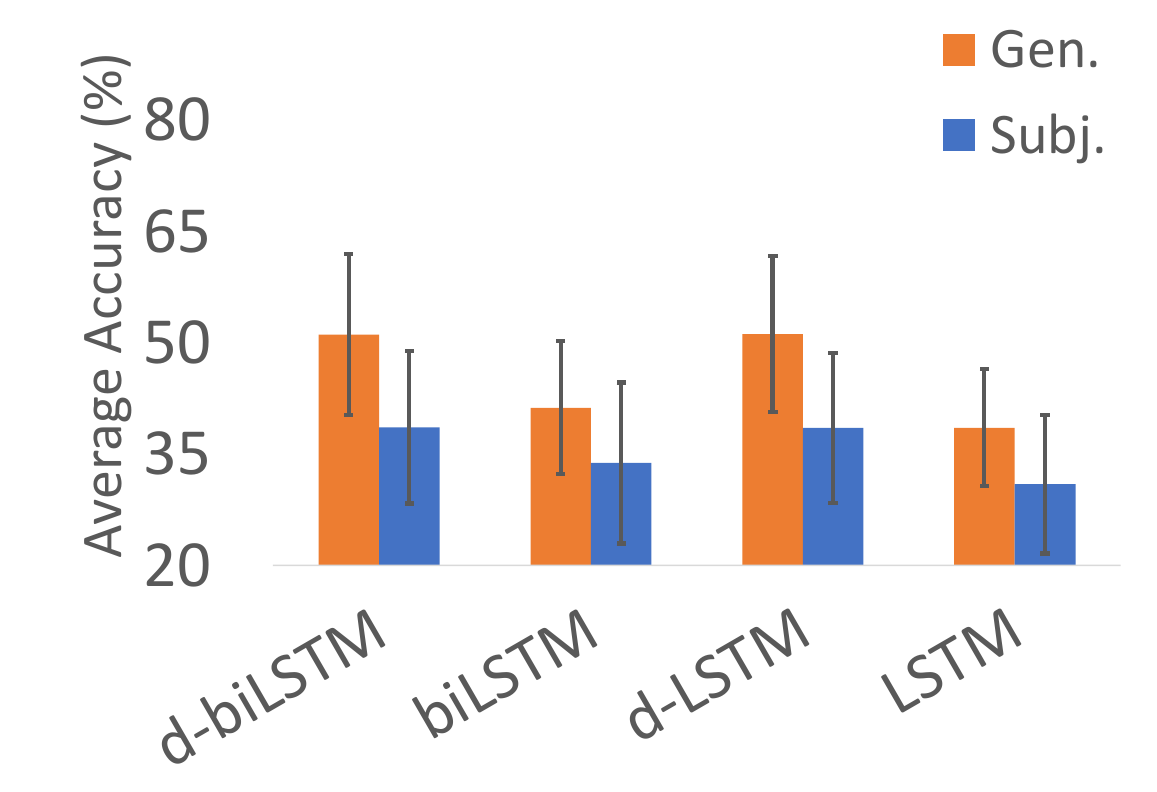}}
    \hfill
	\subfloat[67\% Data]{%
		\includegraphics[width=0.48\columnwidth]{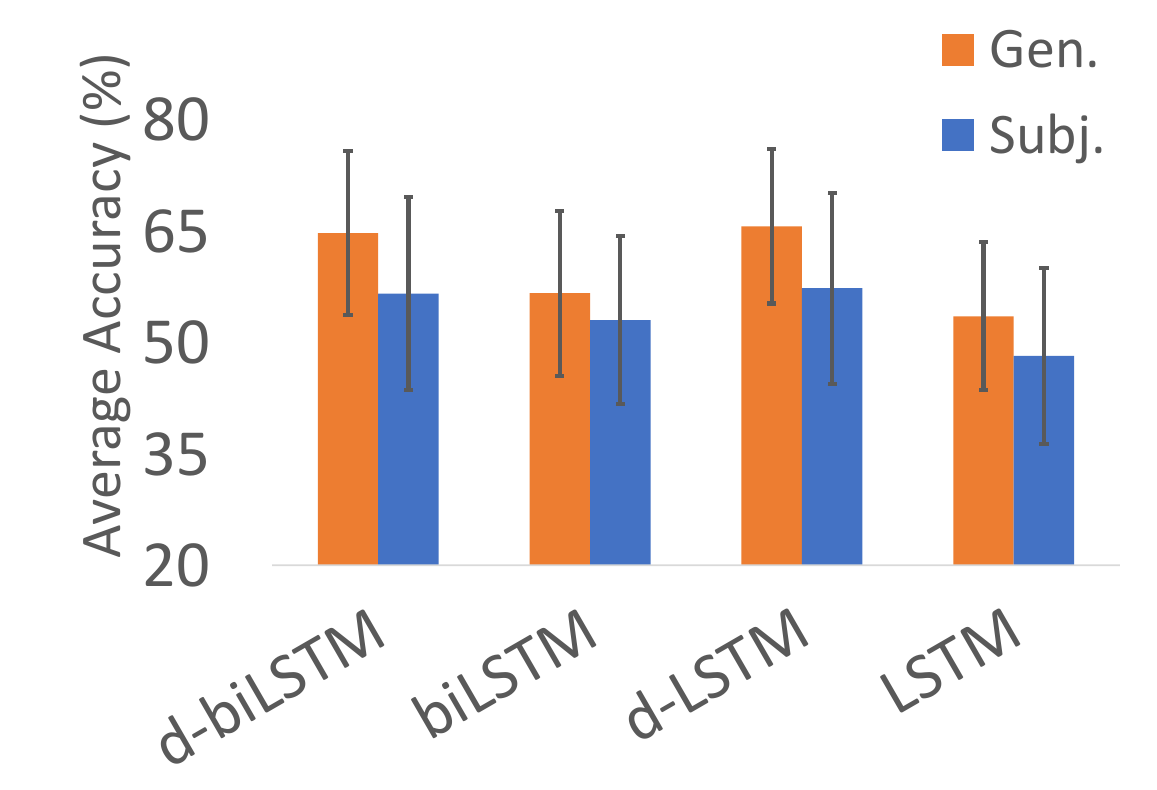}}\\
    \subfloat[100\% Data]{%
		\includegraphics[width=0.48\columnwidth]{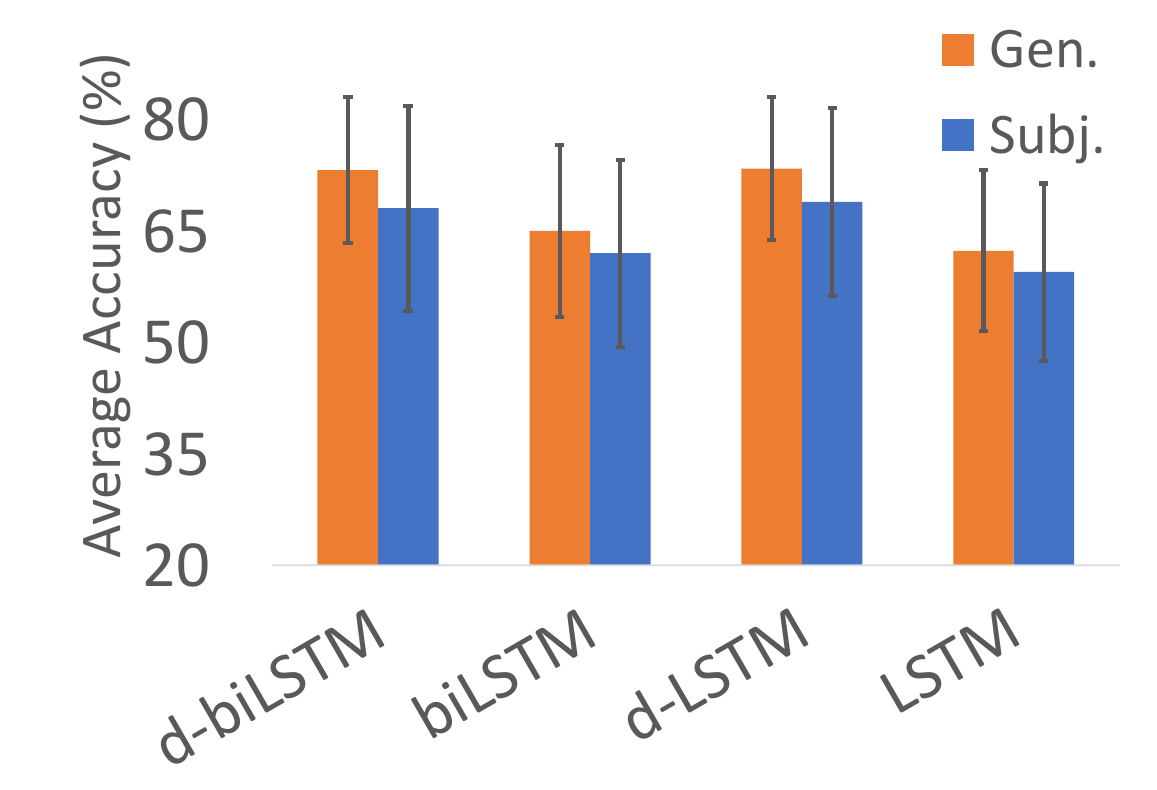}}
    \caption{Comparative results on 65 gestures when given different data availability. The error bars represent one standard deviation away from the mean values.}
	\label{fig:comparative-results}
	\vspace{-1em}
\end{figure}

\begin{figure}[ht]
    \centering
    \includegraphics[width=0.7\columnwidth]{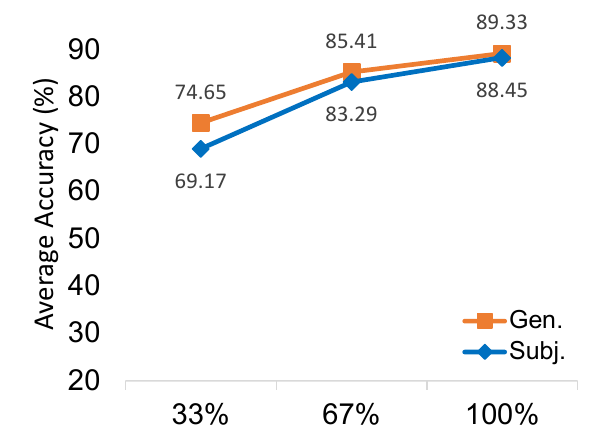}
    \caption{Average results of d-biLSTM on plateau-phase signals given any data availability.} \vspace{-0.3cm}
    \label{fig:comparative_plateau}
\end{figure}

This work presents a novel method for transferring knowledge learned from known subjects to new ones using subject-embedded TL to the best of our knowledge. To demonstrate the superiority of our generalized d-biLSTM model over other state-of-the-art sequential DL models, we apply the same subject-embedded TL to a regular LSTM, a dilated LSTM (d-LSTM), and a regular biLSTM. These models are pre-trained on the same five random subjects (subjects 1, 6, 10, 11, and 14) used by our proposed d-biLSTM. The comparative study is based on 33\%, 67\%, and 100\% of retraining repetitions for all 65 gestures for the remaining 15 subjects. The results are shown in Fig. \ref{fig:comparative-results} and Table \ref{tab:comparative}. It is worth noting that although the d-LSTM model performs similarly to our proposed model, it requires an additional 40,960 trainable parameters (51.9\% more complex than the proposed model) in both generalized and subject-specific models.
In addition, we evaluate the subject-embedded TL technique of our proposed d-biLSTM on signals from the plateau phase (determined as three seconds after the transient phase). The results (shown in Fig. \ref{fig:comparative_plateau}) demonstrate that our proposed method can be applied to signals from any phase of a repetition.

The results demonstrate the impact and potential of our proposed subject-embedded TL approach for generalized muscle activity classification. Analyzing the results of our experiments reveals three main observations: 

\textit{Observation 1:} Dilated models, including the proposed d-biLSTM model, significantly outperform their non-dilated counterparts in terms of classification accuracy across all levels of training data availability. Moreover, the proposed generalized d-biLSTM model exhibits comparable accuracy to the generalized d-LSTM model while requiring fewer trainable parameters, making it a more efficient choice for classification.

\textit{Observation 2:} The generalized models consistently outperform the subject-specific models in both transient and plateau phase experiments for all data availability conditions. This demonstrates the effectiveness of our proposed approach in transferring pre-learned knowledge from known subjects to classify muscle activity from new subjects.

\textit{Observation 3:} The performance gap between the generalized d-biLSTM model and its subject-specific counterpart significantly increases with every 33\% reduction in training data availability. This demonstrates the robustness and generalizability of the proposed approach, even in situations where limited data is available for training.

Overall, these findings suggest that the proposed subject-embedded TL approach using the d-biLSTM model is promising for accurate and efficient muscle activity classification in real-world scenarios involving partially-observed subjects with limited data availability.

\begin{table}
    \renewcommand{\arraystretch}{1.3}
    \caption{Performance comparison between generalized and subject-specific models.}
    \label{tab:comparative}
    \centering
    \begin{tabular}{r||c||c}
        \hline
        \bfseries Model & \bfseries \shortstack{Average Accuracy\\(Generalized)} & \bfseries \shortstack{Average Accuracy\\(Subject-specific)} \\
        \hline\hline
        \textit{\textbf{33\% Data}} \\
        \hline
        d-biLSTM (Ours) & 51.05\% & 38.56\% \\
        Regular biLSTM & 41.22\% & 33.79\% \\
        d-LSTM & 51.14\% & 38.49\% \\
        Regular LSTM & 38.52\% & 30.94\% \\
        \hline
        \textit{\textbf{67\% Data}} \\
        \hline
        d-biLSTM (Ours) & 64.70\% & 56.58\% \\
        Regular biLSTM & 56.62\% & 52.99\% \\
        d-LSTM & 65.62\% & 57.29\% \\
        Regular LSTM & 53.52\% & 48.21\% \\
        \hline
        \textit{\textbf{100\% Data}} \\
        \hline
        d-biLSTM & 73.22\% & 68.04\% \\
        Regular biLSTM & 65.00\% & 62.01\% \\
        d-LSTM & 73.39\% & 68.89\% \\
        Regular LSTM & 62.32\% & 59.47\% \\
        \hline
    \end{tabular}
\end{table}

\section{Conclusions}
\label{sec:conclusions}
In this study, we introduce a subject-embedded TL approach to mitigate the challenge of insufficient training data in DL-based hand gesture recognition. Our proposed d-biLSTM model incorporates a multiplicative embedding layer that encodes subject-specific information, enabling the model to capture subject-specific neurophysiological features while learning HGR pre-knowledge from multiple subjects during pre-training. The resulting generalized models, retrained based on this pre-knowledge, demonstrate superior performance compared to subject-specific counterparts trained from scratch. This performance advantage is particularly evident when data availability is limited, or the number of gestures is large. Additionally, our approach uses transient-phase HD-sEMG signals, corresponding to muscle contraction prior to the maintenance of gestures, to minimize control delay in practical applications. It should be also noted that to the best of our knowledge, our proposed generalized model is the first that includes an embedding layer in a d-biLSTM structure for HGR.

\bibliographystyle{IEEEtran}
\bibliography{references}{}

\vspace{1em}
\vspace{-1.8cm}  \begin{IEEEbiography}
[{\includegraphics[width=1in,height=1.25in,clip,keepaspectratio]{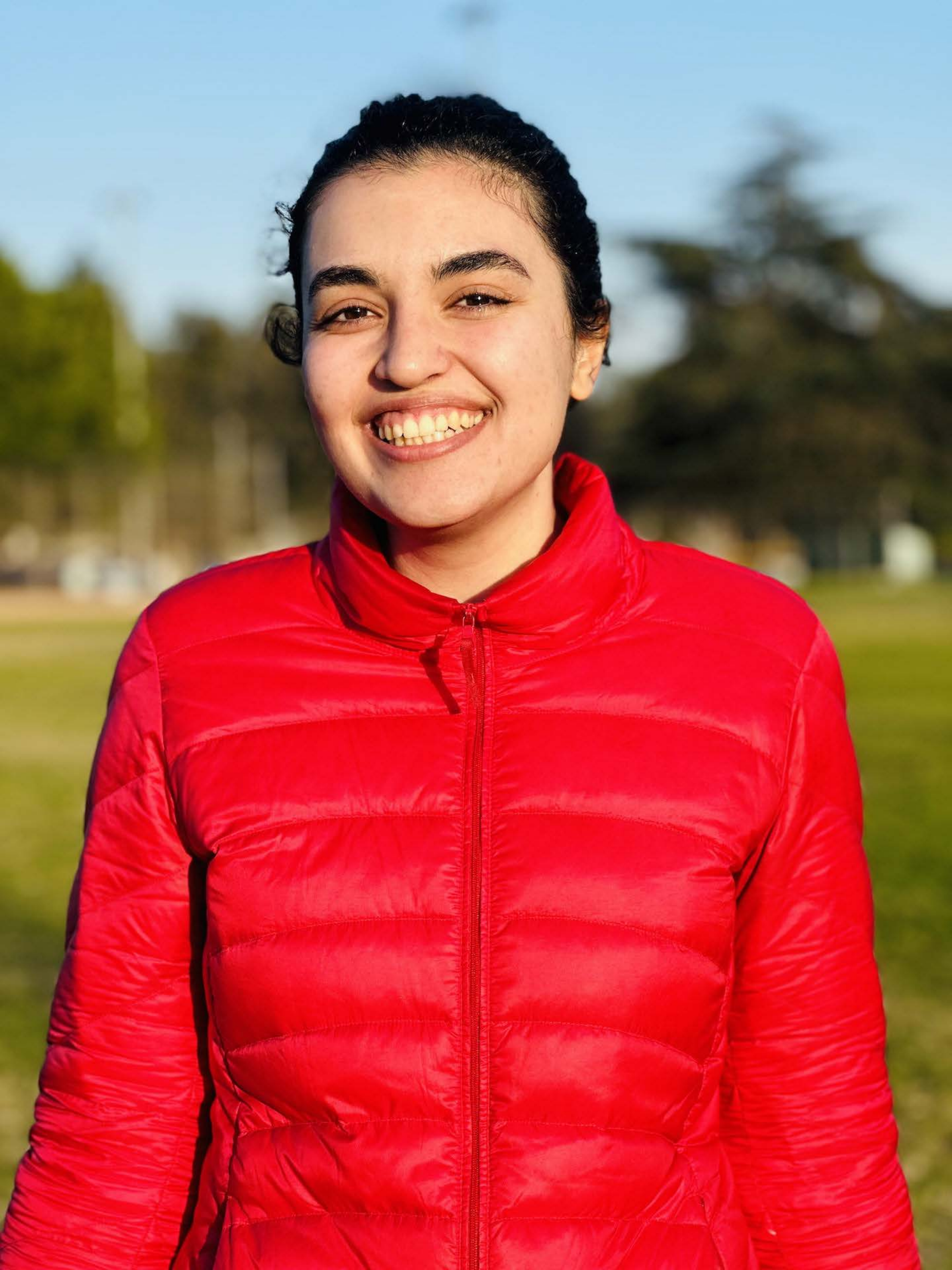}}]{Golara Ahmadi Azar} received her B.S. degree in Electrical Engineering from the Sharif University of Technology and her M.Sc. degree in Electrical and Computer Engineering from the University of California, Los Angeles (UCLA). She is currently a Ph.D. student in the Electrical and Computer Engineering department at UCLA. Her research is focused on theoretical Machine Learning with a focus on asymptotic learning and generalization  with applications in high-dimensional biosignal processing.    \vspace{-1.7cm}
\end{IEEEbiography}
\begin{IEEEbiography}
[{\includegraphics[width=1in,height=1.25in,clip,keepaspectratio]{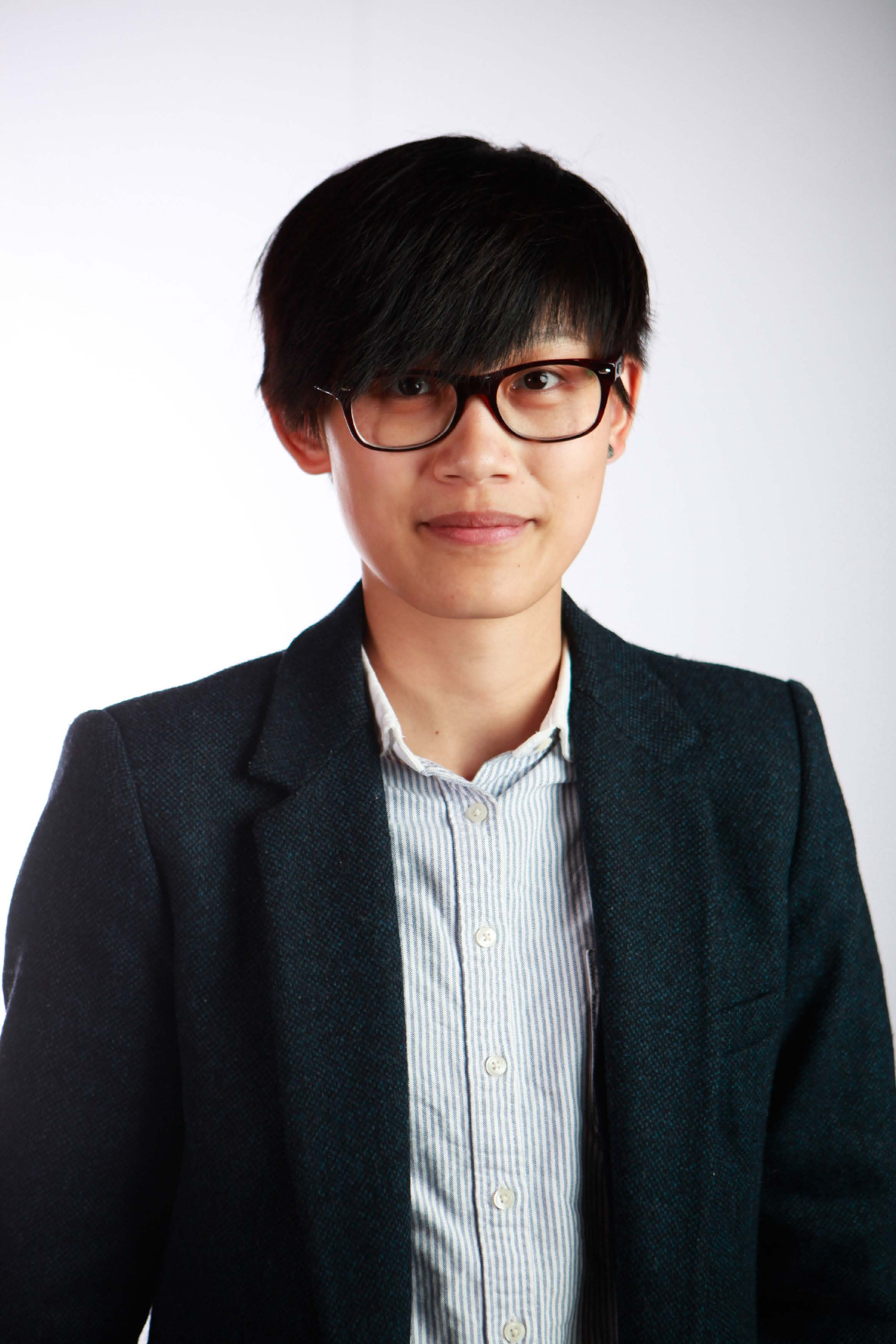}}] {Qin Hu}received her M.Sc. in Quantitative Methods and Modeling from the City University of New York, and M.Sc. in Computer Engineering from New York University (NYU). She is currently a Ph.D. candidate in the Electrical and Computer Engineering department at NYU, working with Medical Robotics and Interactive Intelligent Technologies Laboratory (MERIIT Lab). Her research focuses on explainable, human-centered, and trustworthy AI in biosignal processing, specifically for neural interfaces. \vspace{-2.0cm}  
\end{IEEEbiography}
\begin{IEEEbiography}
[{\includegraphics[width=1in,height=1.25in,clip,keepaspectratio]{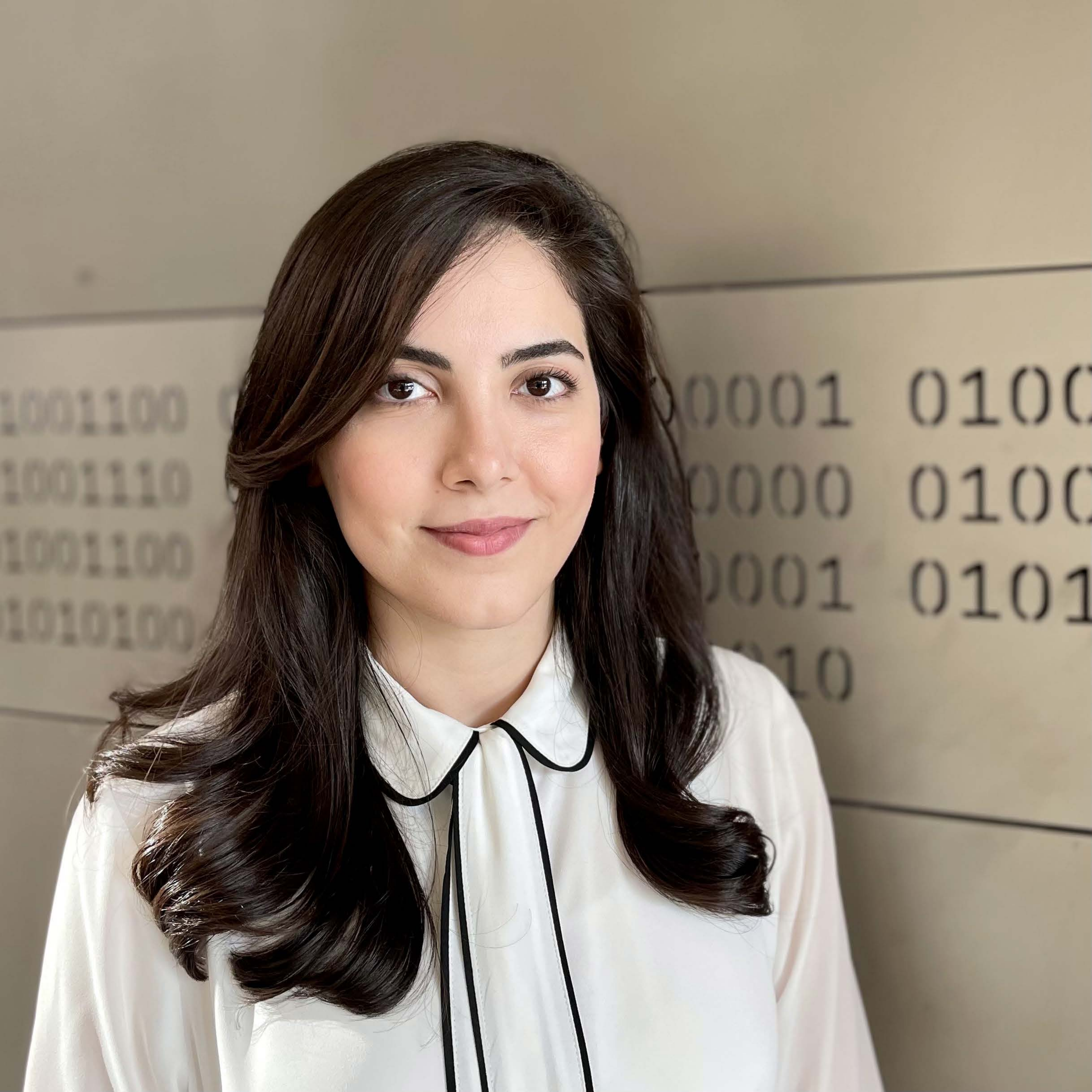}}]{Melika Emami} received her B.Sc. degree in Electrical Engineering from the University of Tehran and M.Sc. and Ph.D. degrees in Electrical and Computer Engineering from the University of California, Los Angeles (UCLA). She is currently a machine learning scientist at Optum AI Labs, working on advancing AI solutions that improve healthcare outcomes.  Her earlier research has been on theoretical machine learning with a focus on the asymptotics for learning and generalization in neural networks.     \vspace{-1.8cm}  
\end{IEEEbiography}
\begin{IEEEbiography}
[{\includegraphics[width=1in,height=1.25in,clip,keepaspectratio]{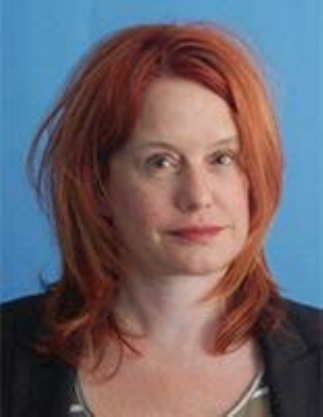}}]{Alyson Fletcher} received the B.S. degree in mathematics from the University of Iowa, Iowa City, IA, USA, and the M.S. degree in mathematics and electrical engineering (EE) and the Ph.D. degree in EE, both from the University of California at Berkeley, Berkeley, CA, USA. Since 2016, she has been on the faculty of the Departments of Statistics, Mathematics, EE, and Computer Science at the University of California, Los Angeles.    \vspace{-1.8cm}  
\end{IEEEbiography}
\begin{IEEEbiography}[{\includegraphics[width=1in,height=1.25in,clip,keepaspectratio]{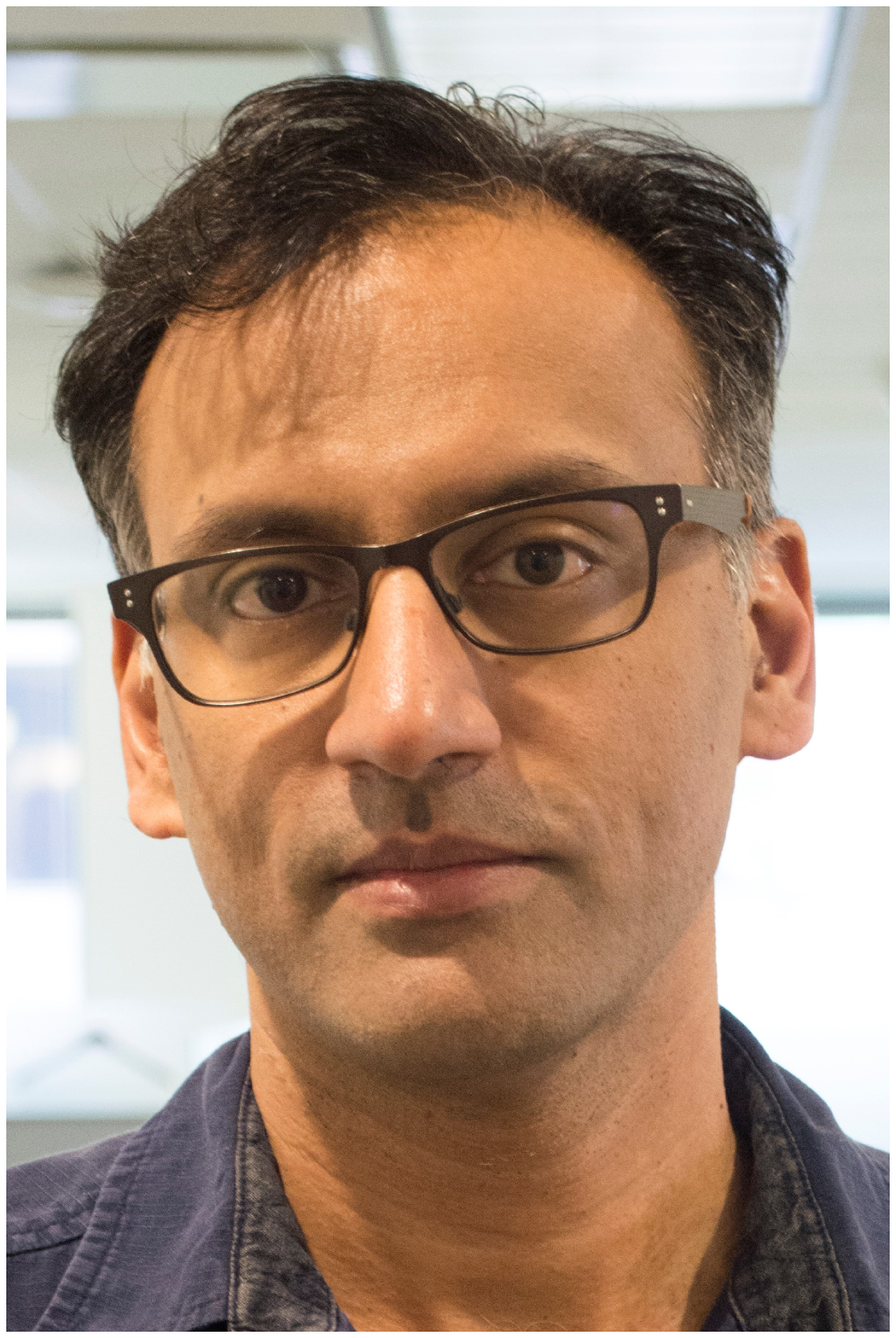}}]{Sundeep Rangan}  received the B.A.Sc. at the University of Waterloo, Canada and the M.Sc. and Ph.D. at the University of California, Berkeley, all in Electrical Engineering. He has held postdoctoral appointments at the University of Michigan, Ann Arbor and Bell Labs.  In 2000, he co-founded (with four others) Flarion Technologies, a spin-off of Bell Labs, that developed Flash OFDM, the first cellular OFDM data system and pre-cursor to 4G cellular systems including LTE and WiMAX. In 2006, Flarion was acquired by Qualcomm Technologies.  Dr. Rangan was a Senior Director of Engineering at Qualcomm involved in OFDM infrastructure products. He joined NYU Tandon (formerly NYU Polytechnic) in 2010 where he is currently a Professor of Electrical and Computer Engineering.  He is a Fellow of the IEEE and the Associate Director of NYU WIRELESS, an industry-academic research center on next-generation wireless systems.
\end{IEEEbiography}
\begin{IEEEbiography}
[{\includegraphics[width=25mm,height=32mm,clip,keepaspectratio]{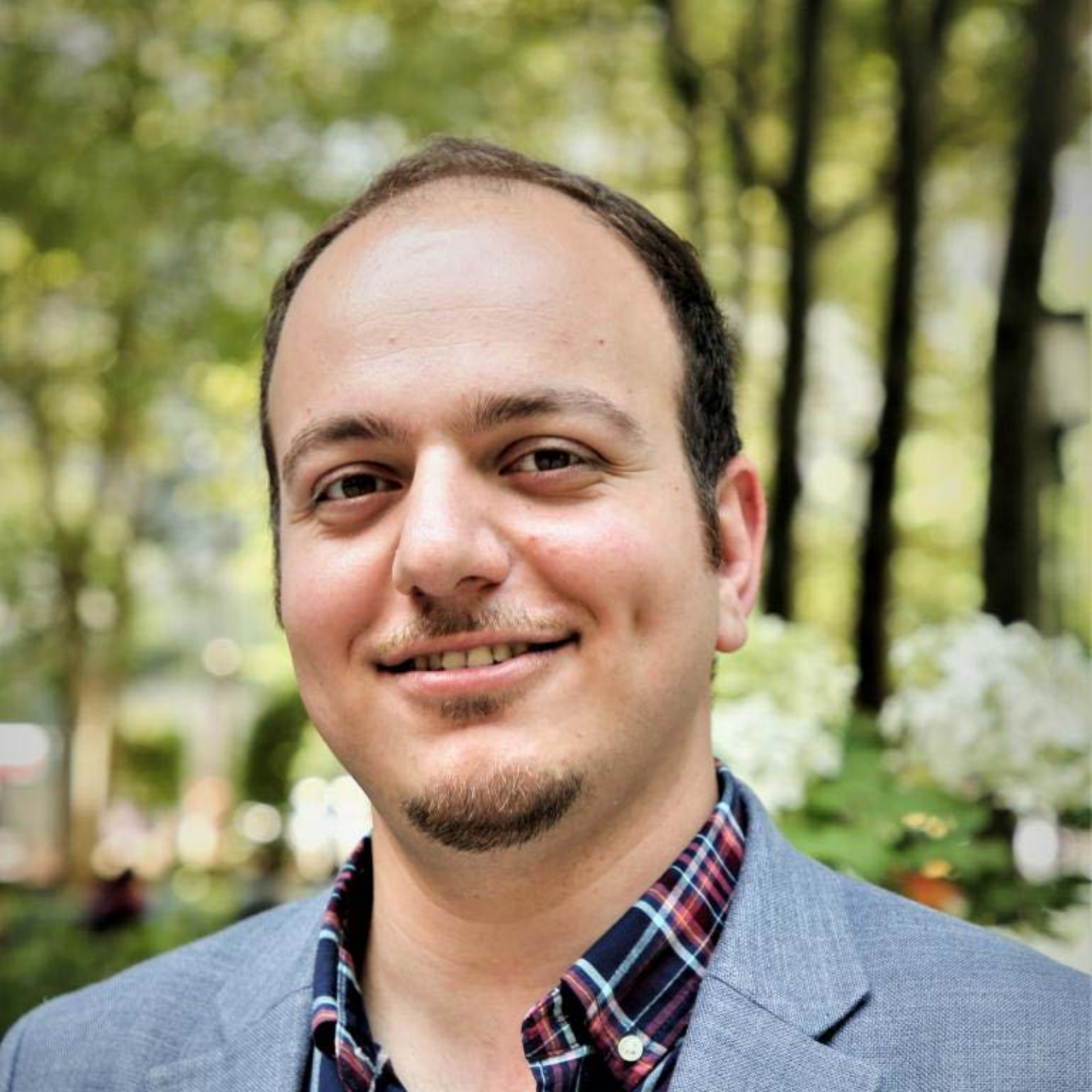}}]{S.Farokh Atashzar}
(Senior Member, IEEE)  is currently an Assistant Professor with New York University (NYU), New York, NY, USA, jointly appointed with the Department of Electrical and Computer Engineering and the Department of Mechanical and Aerospace Engineering. He is also affiliated with the Department of Biomedical Engineering at NYU (New York, NY, USA), also, NYU WIRELESS (New York, NY, USA), and NYU Center for Urban Science and Progress (New York, NY, USA). Before joining NYU, he was a Postdoctoral Scientist at Imperial College London, London, UK. At NYU, he is the director of the Medical Robotics and Interactive Intelligent Technologies (MERIIT) Lab. The lab is mainly funded by the US National Science Foundation. His research interests include a human-machine interface, human-centered robotics, neural interfacing, deep learning, and nonlinear control. Prof. Atashzar was the recipient of several awards, including the 2021 Outstanding Associate Editor of the IEEE Robotics and Robotics. He is currently an Associate Editor for the IEEE Transactions on Robotics and IEEE Robotics and Automation Letters.\vspace{8cm}
\end{IEEEbiography}

\end{document}